\documentclass[sigconf]{acmart}

\AtBeginDocument{%
  \providecommand\BibTeX{{%
    \normalfont B\kern-0.5em{\scshape i\kern-0.25em b}\kern-0.8em\TeX}}}

\newcommand{\tps}{TPS~}
\newcommand{\qq}{Q2Q~}

\usepackage{enumitem}
\usepackage{epsfig,amsmath,amsfonts,multirow,makecell,caption,soul,csquotes,color,wrapfig,subcaption,mathtools,bm,spverbatim,booktabs,xcolor,amsthm,tcolorbox,wrapfig}
\newcounter{countobs}
\usepackage[misc]{ifsym}
\usepackage{bbding}

\copyrightyear{2024}
\acmYear{2024}
\setcopyright{rightsretained}
\acmConference[WWW '24 Companion]{Companion Proceedings of the ACM Web Conference 2024}{May 13--17, 2024}{Singapore, Singapore}
\acmBooktitle{Companion Proceedings of the ACM Web Conference 2024 (WWW '24 Companion), May 13--17, 2024, Singapore, Singapore}\acmDOI{10.1145/3589335.3648335}
\acmISBN{979-8-4007-0172-6/24/05}

\begin{document}

\title{Towards Robustness Analysis of E-Commerce Ranking System}

\author{Ningfei Wang}
\affiliation{%
  \institution{University of California, Irvine}
  \city{Irvine}
  \country{USA}}
\email{ningfei.wang@uci.edu}

\author{Yupin Huang}
\affiliation{%
  \institution{Amazon}
  \city{Palo Alto}
  \country{USA}}
\email{huayupin@amazon.com}

\author{Han Cheng}
\affiliation{%
  \institution{Amazon}
  \city{Palo Alto}
  \country{USA}}
  \email{chenghan@amazon.com}

\author{Jiri Gesi}
\affiliation{%
  \institution{Amazon}
  \city{Palo Alto}
  \country{USA}}
  \email{jirigesi@amazon.com}

\author{Xiaojie Wang}
\affiliation{%
  \institution{Amazon}
  \city{Palo Alto}
  \country{USA}}
  \email{xiojie@amazon.com}


\author{Vivek Mittal}
\affiliation{%
  \institution{Amazon}
  \city{Palo Alto}
  \country{USA}}
  \email{vivekmit@amazon.com}



\begin{abstract}
Information retrieval (IR) is a pivotal component in various applications. Recent advances in machine learning (ML) have enabled the integration of ML algorithms into IR, particularly in ranking systems. While there is a plethora of research on the robustness of ML-based ranking systems, these studies largely neglect commercial e-commerce systems and fail to establish a connection between real-world and manipulated query relevance. In this paper, we present the first systematic measurement study on the robustness of e-commerce ranking systems. We define robustness as the consistency of ranking outcomes for semantically identical queries. To quantitatively analyze robustness, we propose a novel metric that considers both ranking position and item-specific information that are absent in existing metrics. Our large-scale measurement study with real-world data from e-commerce retailers reveals an open opportunity to measure and improve robustness since semantically identical queries often yield inconsistent ranking results. Based on our observations, we propose several solution directions to enhance robustness, such as the use of Large Language Models. Note that the issue of robustness discussed herein does not constitute an error or oversight. Rather, in scenarios where there exists a vast array of choices, it is feasible to present a multitude of products in various permutations, all of which could be equally appealing. However, this extensive selection may lead to customer confusion. As e-commerce retailers use various techniques to improve the quality of search results, we hope that this research offers valuable guidance for measuring the robustness of the ranking systems.
\end{abstract}





\maketitle

\section{Introduction}
\label{sec:intro}
Information retrieval (IR)~\cite{ghorab2013personalised} is a crucial task in various applications such as Web search~\cite{kobayashi2000information}, etc. Unlike other domains, IR is unique in leveraging ranking algorithms to prioritize the relevance of retrieved resources. Therefore, the development of effective ranking systems has consistently occupied a central position in IR. Machine learning (ML) technologies have brought about significant breakthroughs in various longstanding tasks, such as natural language processing (NLP)~\cite{vaswani2017attention}). Therefore, the integration of ML into ranking systems has emerged as a compelling and imperative trend. 

However, ML generally lacks robustness~\cite{li2023sok}, with inherent vulnerabilities to adversarial manipulations~\cite{carlini2017towards, zhang2020interpretable, ma2023slowtrack}. This problem has ignited considerable concerns regarding the robustness and reliability of ranking systems~\cite{song2020adversarial}. Thus, recent research has focused on ranking robustness by generating adversarial examples to evaluate the model's response disparities~\cite{liu2022order, liu2023black, liu2023topic, gupta2021synthesizing}. However, these works have limitations: ignore evaluation in commercialized ranking systems, which is generally more robust than ML models~\cite{wang2023does, shen2022sok, dreossi2019verifai}, and fail to substantiate the relevance between real-world and manipulated queries.
To overcome the first limitation above, we perform the first systematic investigation into the robustness of a leading commercialized e-commerce ranking system. To address the second limitation mentioned above, we use millions of historical search and ranking data sourced from anonymous users in real-world scenarios to assess the system's robustness.

\begin{figure}[!t]
\centering
\includegraphics[width=\linewidth]{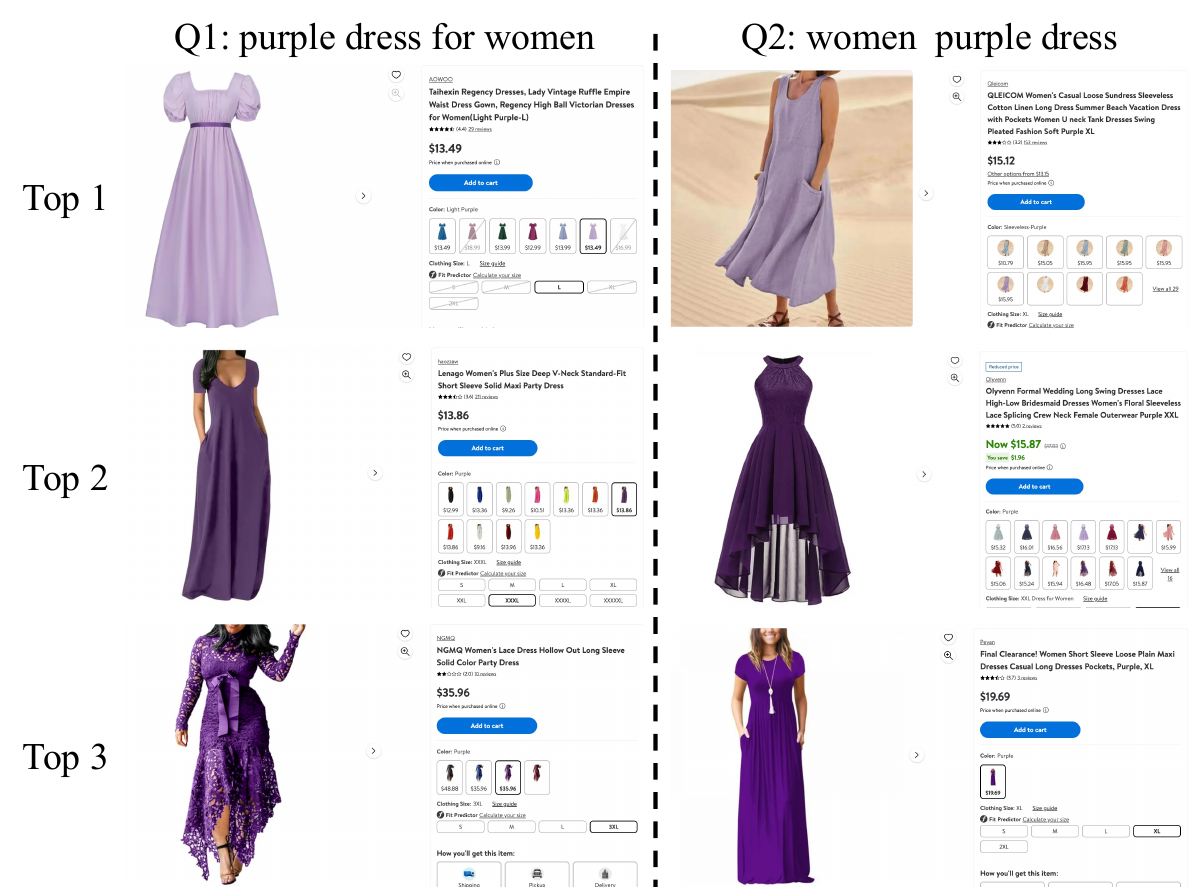}
\vspace{-0.6cm}
\caption{Motivating examples: the two queries are semantically same but their top 3 ranking results are totally different.}
\label{fig:motivating}
\vspace{-0.8cm}
\end{figure}

To initiate the measurement study mentioned above, establishing a formulation for robustness is imperative. Existing formulations highly rely on ground-truth data~\cite{liu2022order, chen2023defense}, which is labeled in the existing dataset by the human judge or approximation from user data. However, in our case, where we leverage real-world data, obtaining reliable and precise ground truth can be challenging or even undesirable, given the vast and diverse range of selections available and nuanced preferences of customers. Therefore, we define ground-truth orthogonal robustness in commercial e-commerce ranking systems as \textit{the consistency of ranking outcomes for semantically identical queries}. This notion gains significance for the following reason: a lack of robustness can severely compromise model accuracy, which is defined as the distance between model prediction and real ground truth~\cite{raghunathan2020understanding}. For instance, low robustness manifests as inconsistent ranking results for semantically identical queries. This inconsistency indicates that the model prediction of one of the queries is likely to exhibit a significant departure from the ground truth. Such non-robust behavior has the potential to undermine user experience and thus reduce the customer's confidence in trusting the ranking results from the e-commerce systems. 

To quantitatively analyze the robustness defined above, an appropriate metric is necessary. With a comprehensive review of the prior metrics, we have identified noteworthy shortcomings in all existing metrics. They tend to equally penalize errors occurring at any position within the ranking list and struggle to handle cases where items appear in one list but not the other. With these, we propose a novel metric to assess the disparity between two ranking lists, considering both the position and item-specific information.

We conduct the first large-scale measurement study on the ranking system in one of the leading e-commerce retailers, with our newly proposed metric and historical real-world data. We study more than several million queries and group them together as query pairs by their semantics with rule-based normalization techniques~\cite{bialecki2012apache} and predictive modeling method~\cite{Huang2023}. An example is shown in Fig.~\ref{fig:motivating}, where we analyze a pair of semantically identical queries: `purple dress for women' and `women purple dress'. Notably, the top 3 ranking results were entirely different. From a model training perspective, this finding implies that at least one query is likely to yield rankings divergent from the ground truth due to semantic consistency but ranking inconsistency. Our large-scale study reveals new and open research opportunities to improve robustness of ranking system, as it frequently generates divergent ranking outcomes for semantically identical queries. The issue of robustness discussed herein does not constitute an error or oversight. Instead, in situations characterized by a wide range of options, it is possible to offer multiple products in various configurations, each appealing to different customers. Nevertheless, this abundance of choices may cause consumer confusion. 

To ensure the semantic identity of our evaluated query pairs, we conducted a user study. The results of this evaluation indicate that a significant majority of the query pairs analyzed in our study are indeed perceived as semantically identical by human participants. In order to systematically explore the robustness of the ranking system, we also present a comprehensive taxonomy delineating various patterns observed in non-robust query pairs, the temporal evolution of robustness, and the integration of robustness into commercialized models. More details are demonstrated in~\S\ref{sec:eval}.

Building upon the insights garnered from our measurement study, we propose several solutions and outline future research directions to improve ranking system robustness. Specifically, we investigate the potential of Large Language Models (LLMs) and model ensembles as avenues for enhancements in the robustness of e-commerce ranking systems. 
We hope that the insights will serve as valuable guidance for the design of e-commerce ranking systems, facilitating enhancements in both their robustness and accuracy.

To sum up, this paper makes the following contributions:
\begin{itemize}[leftmargin=*]
  \item We conduct the first large-scale measurement study on the robustness analysis of leading commercialized e-commerce ranking system, utilizing millions of real-world data from users.
  \item We propose a novel metric designed to quantitatively assess the disparity between two ranking lists for the ranking systems.
  \item We systematically measure the robustness of the e-commerce ranking system by addressing five research questions. Additionally, we present ten measurement observations that could guide future robustness and accuracy improvement of ranking systems.
\end{itemize}
\section{Related Work}
\label{sec:literature}
\textbf{Text-based model robustness.}
Recent works find that ML models are generally not robust in different NLP tasks~\cite{zeng2020openattack, morris2020textattack, simoncini2021seqattack, xu2020adversarial, ye2022texthoaxer, yuan2021bridge, maheshwary2021strong}.
In response, a plethora of studies have endeavored to enhance their robustness, such as adversarial training~\cite{ivgi2021achieving, zhu2019freelb, wang2021adversarial, liu2020robust} and certified robustness~\cite{wang2021certified, xu2020automatic, ye2020safer, shi2020robustness, huang2019achieving, jia2019certified, ko2019popqorn}. Given that ranking systems share the ML nature, they inherit similar robustness challenges, which can impact user experiences and harm a company's reputation. However, currently, there is a limited number of research studies dedicated to investigating the robustness issues of ranking systems~\cite{liu2022order, chen2023defense}, especially on the robustness improvement. Thus, we conduct the first large-scale measurement study to assess the robustness of commercialized e-commerce ranking systems.

\textbf{Ranking metrics.}
To evaluate the performance of ranking models, a wide range of metrics, e.g., normalized discounted cumulative gain (NDCG)~\cite{jarvelin2002cumulated, wang2013theoretical} and expected reciprocal rank (ERR)~\cite{chapelle2009expected}, have been introduced~\cite{wang2016evaluating,wang2017search}. However, these conventional metrics rely on ground-truth data, making them unsuitable for the problem settings in this paper. Furthermore, metrics such as Kendall’s $\tau$~\cite{kendall1938new}, $\tau_{AP}$~\cite{yilmaz2008new}, $\tau_{a}$~\cite{urbano2017treatment}, $\tau_{b}$~\cite{urbano2017treatment}, and Spearman's rank correlation coefficient~\cite{sedgwick2014spearman} have been developed to quantify the disparity between two ranking lists~\cite{wang2023well}. However, these existing metrics exhibit limitations, including the equal penalization of errors across all positions in the list and inadequate handling of items that appear in one list but are absent in the other. We provide detailed discussion in~\S\ref{sec:design}. Based on their drawbacks, we introduce a novel metric specifically tailored to evaluate the robustness of ranking systems in e-commerce settings, which will be detailed in~\S\ref{sec:design}.

\section{Methodology}
\label{sec:design}

\subsection{Ranking Model}
\label{sec:rankingmodel}
Ranking models are essential for ordering candidate items by relevance to a query. Traditional methods such as BM25~\cite{robertson1994some} struggle to model human language effectively. Recent advancements in ML~\cite{vaswani2017attention} have led to ML-based models emerging as state-of-the-art solutions for various ranking tasks\cite{chu2022h, dehghani2017neural, zamani2018neural}. Nonetheless, these ML-based ranking models exhibit robustness issues~\cite{liu2022order, liu2023topic, liu2023black, wang10bert, wu2023prada}.
Notably, the literature has not thoroughly evaluated the robustness of commercial e-commerce ranking systems or delineated the distinction between real-world and manipulated queries. Commercial ranking systems, such as Google or Amazon, often exhibit greater robustness than standalone ML models~\cite{wang2023does, shen2022sok, dreossi2019verifai}. For example, these systems usually include query rewrite, auto-completion, and search facet to create a smooth search journey, which can potentially avoid the robustness problem from tweaking characters~\cite{liu2022character} in the input. Thus, it is unclear how robust the commercial e-commerce ranking models are with real-world queries. Thereby, we perform the first measurement study on the commercial e-commerce ranking system robustness with real-world queries to fill in these research gaps.

We provide the formulation of the ranking system. Given a query $q$, the ranking systems generate a list of ranked items, denoted as <$i_1, i_2, \cdots, i_n$>, along with their respective positions in the list, denoted as <$p_1, p_2, \cdots, p_n$>. Typically, these positions are determined based on the relevance scores predicted by the ranking models~\cite{liu2022order}. These results are subsequently presented to users through systems.

\subsection{Robustness Scope and Motivating Example}
\label{sec:formulation}

{\bf Robustness scope.} We define robustness in e-commerce ranking systems as the \textit{consistency of ranking results between two semantically identical queries}. Under an ideal scenario with a ranking system that is both robust and accurate, two distinct but semantically equivalent queries would yield identical ranking outcomes. Conversely, a system that produces different ranking results for such queries—despite all items being relevant—is considered non-robust. Such a problem is critical considering the following scenarios. If ground truth for the ranking model is available, the lack of robustness can compromise the model's accuracy when compared to the ground truth, owing to the inconsistency in ranking outcomes; inevitably, one of the queries is likely to deviate significantly from the ground truth. Moreover, considering that e-commerce models are generally trained on user behavior data, non-robust outcomes generated by the model could induce users to engage in suboptimal behaviors. This, in turn, could result in biased data, thereby affecting the subsequent training and performance of the model.

\label{sec:motivate-example}
{\bf Motivating example.} We consider the two semantically identical queries: $Q_1$ as `purple dress for women' and $Q_2$ as `women purple dress' in Fig.~\ref{fig:motivating}. When searching on one of the leading commercialized e-commerce retailers, the ranking lists for these queries were completely dissimilar. Fig.~\ref{fig:motivating} shows a comparison of the top 3 ranking results for both queries.
This stark contrast in ranking outcomes for semantically equivalent queries suggests that there is at least one query ($Q_1$ or $Q_2$) for which the model predictions differ from the ground truth, despite the queries being semantically equivalent. This observation underscores the fundamental issue studied in this paper, which we refer to as model robustness.

\subsection{Robustness Study Method}
\label{sec:robust_method}
To address the robustness problem in~\S\ref{sec:formulation}, the initial and indispensable step is to identify queries that contain different forms or alternative expressions of the same concept. We explore two types of data in this paper: 1) Rule-based normalization query pair, and 2) Predictive modeling query pair.

\textbf{Rule-based Normalization Data.} We leverage query normalization techniques~\cite{manning2008} that include tokenization, filtering, and stemming to consolidate the queries. Further, we observe that shopping intent queries are typically concise. Reordering the tokens within these queries usually does not alter the underlying concept. Thus, in Text Processing and Sorting (TPS) data, queries sharing the same normalized tokens are considered as semantically identical query pairs, such as ``battery AA'' and ``AA battery''.

\textbf{Predictive Modeling Data.} To improve the behavior feature with query cluster and feature coverage by mapping queries to the closest query, \qq (in-house trained query to query similarity prediction) models are proposed with collected similar queries and created labels~\cite{Huang2023}. 
\qq can provide similarity scores between the two query pairs, and thus, can be used in our measurement study.

For 1), our approach begins by collecting vast data of query pairs. With that, we select a subset in which the queries exhibit the same TPS. The selection process is crucial, as it ensures that the retained query pairs share a meaningful level of semantic equivalence. In 2), we adopt the following procedure: one query designated as query 1 is held constant, and we then identify the top k query 2 candidates with the highest similarity scores to create query pairs. Then, we employ these query pairs to retrieve historical data from e-commerce systems. This process allows us to identify the ranked items with their unique identifiers and average ranking positions.

\vspace{-0.2cm}
\subsection{Metric Design}
\label{sec:metric}
As introduced in~\S\ref{sec:literature}, most representative metric designs such as Kendall's $\tau$ and $\tau_{AP}$ suffer from several drawbacks within the context of our study: 1) equally penalizing errors that occur at any position in the list, which fails to fully consider the order information; 2) difficulty in handling cases where items appear in one list but not in the other. Kendall's $\tau$ is particularly deficient in both aspects, while $\tau_{AP}$ is mainly affected by the latter issue.

For instance, regarding the first drawback 1), consider three ranking lists: R1: <1, 2, 3, 4>, R2: <1, 2, 4, 3>, and R3: <2, 1, 3, 4>. Given the significance of item position in ranking lists, R1 and R2 should intuitively be more similar than R1 and R3. However, when calculating Kendall’s $\tau$, we obtain the same result of 0.67 for both comparisons, highlighting the limitation in addressing this first drawback. In the context of the second drawback 2), we examine another three ranking lists: R1: <1, 2, 3, 4>, R2: <1, 2, 3, 4>, and R3: <1, 2, 5, 6>. R1 and R2 should be more similar than R1 and R3 since R1 and R2 are exactly the same. However, both Kendall’s $\tau$ and $\tau_{AP}$ yield values of 1.00 for these comparisons. This experimental analysis underscores the limitations of the representative existing metric in handling this second drawback. These limitations are not unique to these two metrics but also affect other state-of-the-art ranking metrics, such as Spearman's rank correlation coefficient~\cite{sedgwick2014spearman}. These limitations hinder the applicability of these metrics in our context.

Addressing the second limitation requires a method to account for missing items in ranked lists. One straightforward approach is to append these missing elements to the end of the list. Integrating this method with specific metrics, such as $\tau_{AP}$, may offer a solution to both identified defects. However, this direct appending method introduces unique corner cases, wherein the appended items create discrepancies in position that can result in counterintuitive errors. Consider the evaluation of three distinct ranking lists: R1: <1, 2, 3, 4>, R2: <1, 2, 4, 3>, and R3: <1, 2, 5, 6>. The $\tau_{AP}$ value stands at 0.33 for R1 and R2. When we apply the appending strategy to R1 and R3, the process entails calculating the $\tau_{AP}$ between the lists <1, 2, 3, 4, 5, 6> and <1, 2, 5, 6, 3, 4>. Notably, this also yields a $\tau_{AP}$ of 0.33, identical to the value derived from comparing R1 and R2. Intuitively, the similarity between R1 and R2 should be larger than that between R1 and R3. Thus, this equivalence in $\tau_{AP}$ values underscores an inherent flaw in the appending approach. Thereby, the weakness of this straightforward strategy renders it unsuitable for addressing the nuances of our problem context.

With limitations observed in existing metrics, we propose a novel metric named RDS (Ranking Distance Score) defined in Eq (\ref{eq:metric}) that addresses these issues. When comparing two ranking lists, $R_1$ and $R_2$, RDS incorporates a position-based decay using a logarithmic function and also checks for the presence of items in both lists. 
\begin{equation}
\label{eq:metric}
     \sum_{r \in R_1 \cup R_2} RDS (p(r, R_1), p(r, R_2))\\
\end{equation}
where the function $p(*)$ is to retrieve the position of an item in the ranking list and returns -1 if the item does not exist in the ranking list. The function $RDS(*)$ is to calculate the score defined in Eq (\ref{eq:metric-if}).
\begin{equation}
\label{eq:metric-if}
   RDS (p_1, p_2)=  \left \{
    	\begin{array}{l}
    		\lvert \frac{1}{\log_2 (p_1 + 1)} - \frac{1}{\log_2 (p_2 + 1)}\rvert, \ p_1 \neq -1 \ \land \ p_2 \neq -1 \\
    		\lvert \frac{1}{\log_2 (2)} - \frac{1}{\log_2 (p_{1_{max}} + 1)}\rvert + \frac{1}{\log_2 (p_1 + 1)}, \ p_1 \neq -1 \\
          \lvert \frac{1}{\log_2 (2)} - \frac{1}{\log_2 (p_{2_{max}} + 1)}\rvert + \frac{1}{\log_2 (p_2 + 1)}, \ p_2 \neq -1
    	\end{array}
	\right.
\end{equation}
where the $p_{*_{max}}$ represents the maximum position in the ranking list $*$ (either 1 or 2). The term $\lvert \frac{1}{\log_2 (2)} - \frac{1}{\log_2 (p_{*_{max}} + 1)}\rvert$ is to address the second drawback, which is as a penalty term to overcome the limitation of handling items that appear in one list but not the other.

RDS can effectively address the limitations mentioned above
For instance, consider three ranking lists: R1: <1, 2, 3, 4>, R2: <1, 2, 3, 4>, and R3: <1, 2, 5, 6>. After normalizing the scores to a range of 0 to 1 and calculating the similarity score (i.e., 1 minus the RDS), we find that the score between R1 and R2 is 1.00, while the score between R1 and R3 is 0.38. For the case: R1: <1, 2, 3, 4>, R2: <1, 2, 4, 3>, and R3: <1, 2, 5, 6>, the similarity score between R1 and R2 is 0.95, while the similarity score between R1 and R3 is 0.38. These simulated results clearly demonstrate that our metric outperforms existing ones even the existing metric with a straightforward design in effectively capturing the differences in ranking list similarity.
\section{Measurement Study}
\label{sec:eval}
\setcounter{countobs}{0}

\subsection{Methodology and Setup}
\label{sec:methodandsetup}
{\bf Data preparation.} This study utilizes two distinct datasets: 1) a collection of real-world historical search and ranking data retrieved weekly from the ranking system, and 2) a dataset derived from the in-house \qq model. The first dataset comprises several billion data points that facilitate the generation of \tps data pairs, as introduced in \S\ref{sec:robust_method}, and aid in constructing evaluation sets for ranking lists. These lists comprise items and their corresponding average positions given a specific query. Within this context, we have generated over 26 million query pairs, their relevance determined based on whether each pair shares identical TPS. The second dataset, encompassing roughly several billion query pairs, allows for the generation of more than 4 million query pairs. This process is achieved by holding one query constant and pairing it with the top k queries based on the highest similarity scores within the \qq model. We assume that both \tps and \qq model similarity scores can offer a semantically identical guarantee to a certain degree, which will be further explored in \S\ref{sec:rq2}. All data used in this study is sourced from real-world users, lending a high degree of real-world applicability to our findings. Note that all the data is anonymous, and no other potentially private data is included.

\textbf{Data filter.} 
To enhance the quality of the data, we apply a series of filters to eliminate noise. First, we restrict our data originating from the United States and written in English to control regional and linguistic biases. Second, we eliminate the bottom 20\% of query-item pairs based on their searched frequency from users within the week to remove low-frequency errors that could affect ranking. Third, we consider only those cases where the ranking list length exceeds 20, specifically the top 20 ranking results. Last, for query lists sharing the same TPS, we include only the top three queries based on weekly search count to control the bias of query frequency.

\textbf{Large-scale measurement study.} 
We perform our large-scale evaluation using PySpark~\cite{drabas2017learning} on the Amazon EMR cluster. The study focuses on the e-commerce system, generating ranked lists from historical real-world anonymous user query data.

\subsection{Research Question}
We first pose several research questions (RQs) to assess the robustness and semantic consistency of e-commerce ranking systems:

\begin{itemize}[leftmargin=0.4cm]
  \item RQ1: How robust is the e-commerce ranking system?
  \item RQ2: To what extent can semantic consistency be ensured?
  \item RQ3: How has the robustness of the e-commerce ranking system evolved over time?
  \item RQ4: How is robustness incorporated into the in-house Q2Q model used for the e-commerce ranking system?
  \item RQ5: What types of query pairs challenge the robustness of the e-commerce ranking system?
\end{itemize}

\subsection{RQ1: Robustness of E-Commerce System}
\label{sec:rq1}
\begin{figure}[!t]
\centering
\includegraphics[width=\linewidth]{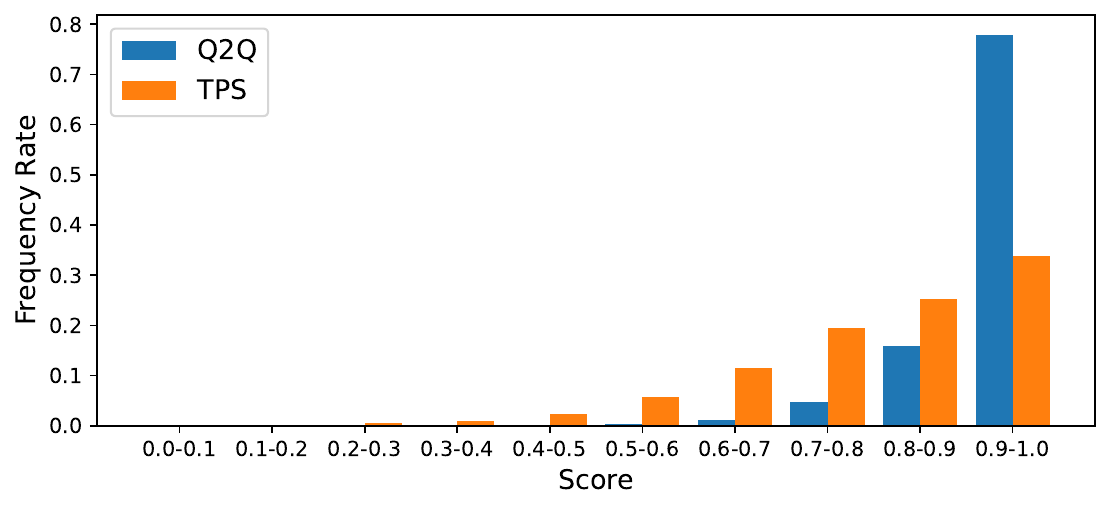}
\vspace{-0.8cm}
\caption{Histogram of \tps and \qq data evaluated with RDS metric on millions of query pairs.}
\label{fig:tps_q2q}
\vspace{-0.3cm}
\end{figure}

We employ RDS metrics to evaluate the robustness of the e-commerce ranking system, utilizing both \tps and \qq data types described in~\S\ref{sec:methodandsetup}. Our findings, illustrated in Fig.~\ref{fig:tps_q2q}, reveal substantial discrepancies in the ranking outcomes provided by the e-commerce ranking system for most query pairs in both data types. Notably, the \qq data exhibits a frequency rate for RDS between 0.9 and 1.0 that is more than twice as high as that in the \tps data. This suggests that the \qq data imposes fewer constraints on the semantics of the query pairs, making it less robust compared to TPS. Thus, in the following analysis, we mostly use \tps data since it can help to understand how robust the current e-commerce system is with tight constraints on the semantics.
With that, we can conclude:
\begin{tcolorbox}[boxrule=0pt, left = 2mm, right = 2mm, top = 0.5mm, bottom = 0.5mm, title= Observation \stepcounter{countobs}\arabic{countobs}]
The current commercialized e-commerce ranking system displays a lack of robustness as it often produces divergent ranking outcomes for semantically identical queries.
\end{tcolorbox}

\subsection{RQ2: Semantic Consistency Guarantee}
\label{sec:rq2}
To systematically understand whether the two queries are semantically identical, we perform a user study.

\begin{figure}[!t]
\centering
\includegraphics[width=\linewidth]{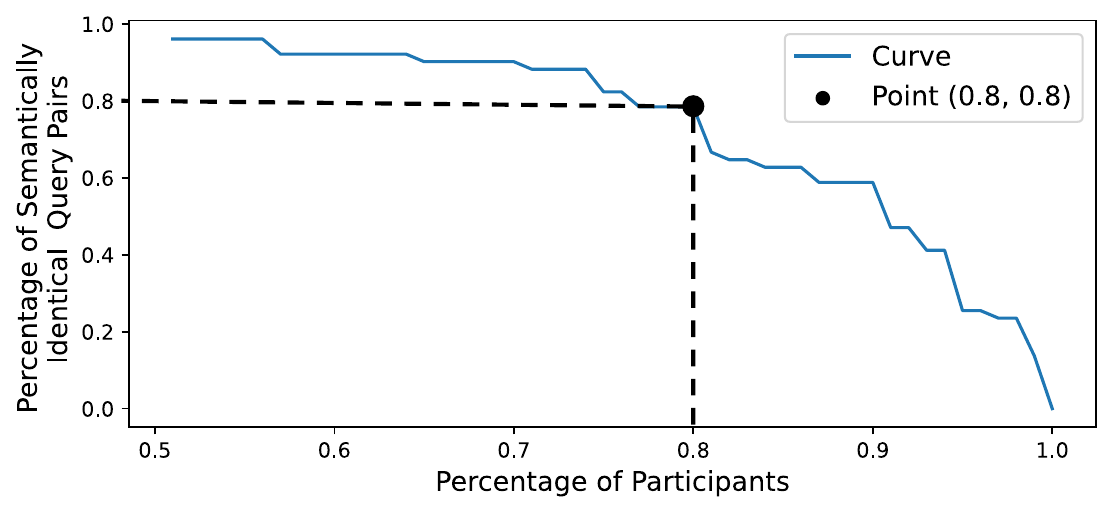}
\vspace{-0.8cm}
\caption{Illustration of participants and semantically identical query pairs for user study. Specifically, 80\% of participants regarded 80\% of the query pairs as semantically identical.}
\label{fig:cdf}
\vspace{-0.5cm}
\end{figure}

\textbf{Methodology and setup.} We recruited 50 human subjects through Prolific~\cite{Prolific}, a platform specializing in research-related crowd-sourcing. All participants, ranging in age from 19 to 72, were verified to have experience shopping on one of the leading e-commerce retailers and proficiency in English. Each subject was presented with 50 sets of query pairs, randomly sampled from our full \tps data. We describe the shopping scenarios and ask the survey question: ``in a hypothetical shopping scenario online, one might input two semantically identical queries into the search bar with the expectation of receiving identical search results. We hope that you can help to judge whether you are expecting to get the same results from e-commerce retailers giving two queries.'' Then, we provide different groups of query pairs and ask the users to provide binary answers, i.e., ``Yes'' or ``No''. 

\textbf{Results.} 
As depicted in Fig.~\ref{fig:cdf}, our survey indicates that 80\% of the participants perceive 80\% of the query pairs to be semantically identical and expected e-commerce retailers to return identical results for these query pairs. This finding corroborates our initial hypothesis regarding semantic similarity among the query pairs. Based on the results, we can formulate the following observation:

\begin{tcolorbox}[boxrule=0pt, left = 2mm, right = 2mm, top = 0.5mm, bottom = 0.5mm, title= Observation \stepcounter{countobs}\arabic{countobs}]
The majority of query pairs are perceived as semantically identical by human subjects, strengthening the observation of a lack of robustness in the e-commerce ranking system.
\end{tcolorbox}
 
\subsection{RQ3: Evolution of Robustness}
\label{sec:rq3}
\begin{figure}[!t]
\centering
\includegraphics[width=\linewidth]{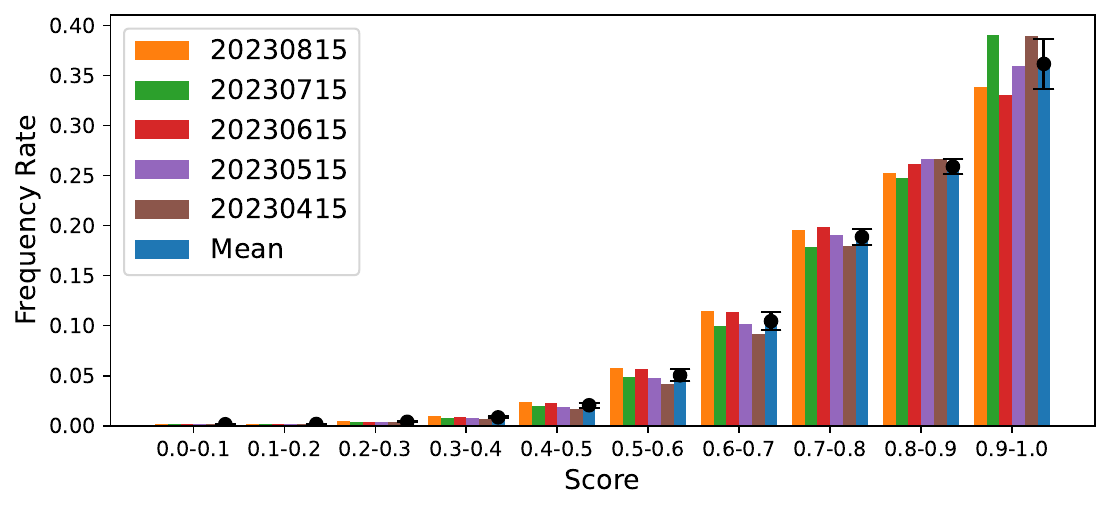}
\vspace{-0.8cm}
\caption{Histogram for RDS from \tps data over a five-month period with more than 20 million query pairs for each time.}
\label{fig:tps_5weeks}
\vspace{-0.5cm}
\end{figure}

To address RQ3, we assess the evolution of e-commerce ranking system robustness over a period of four months, spanning from April 15, 2023, to August 15, 2023. During this period, we conduct evaluations on a weekly basis for one selected week each month, and the results are presented in Fig.~\ref{fig:tps_5weeks}. We use \tps data for analysis.

In Fig.~\ref{fig:tps_5weeks}, we also report key statistical measures, including the average score and the standard deviation (STD). The figure includes a detailed histogram illustrating these statistics. The frequency rate difference across bins is observed to be minimal. Thus, the STD values are notably low. Specifically, the maximum STD observed is 0.02, which is associated with an average normalized frequency rate of 0.35. Given the small magnitude of the STD, we conclude that the variance is negligible in comparison to the original values.

\begin{tcolorbox}[boxrule=0pt, left = 2mm, right = 2mm, top = 0.5mm, bottom = 0.5mm, title= Observation \stepcounter{countobs}\arabic{countobs}]
Our analysis indicates that the e-commerce system's robustness has remained stable over the observed period, suggesting that the robustness has not been considered to be improved.
\end{tcolorbox}

\subsection{RQ4: Robustness Integration on \qq}
\label{sec:rq4}
To investigate RQ4, we examine the relationship between the similarity scores utilized in in-house \qq models~\cite{Huang2023} and our RDS metric. We use the \textit{Pearson correlation coefficient} metrics, as implemented in the Scipy Python library~\cite{2020SciPy-NMeth}.

Our results indicate that the absolute value of the \textit{Pearson correlation coefficient} is 0.345, thereby indicating a low correlation between the two metrics~\cite{mukaka2012guide}. Additionally, a \textit{p-value} of 0.0 confirms that the observed correlation is statistically significant~\cite{dahiru2008p}. Given that in-house Q2Q models are trained and evaluated based on similarity scores~\cite{Huang2023}, it is evident that robustness considerations are largely absent from both the training and evaluation processes.

\begin{tcolorbox}[boxrule=0pt, left = 2mm, right = 2mm, top = 0.5mm, bottom = 0.5mm, title= Observation \stepcounter{countobs}\arabic{countobs}]
Our findings suggest that robustness factors are generally not incorporated into the training or evaluation of in-house \qq models in e-commerce ranking systems.
\end{tcolorbox}

\subsection{RQ5: Taxonomy of Non-Robust Cases}
\label{sec:rq5}
Addressing RQ5 is pivotal for systematically improving the robustness of current e-commerce ranking systems. To this end, we conduct a novel classification study using historical data from real users, focusing on the worst cases of which RDS is 1.0.

\textbf{Taxonomy methodology.}
Due to large amounts of \tps query pairs, i.e., more than 20 million, it is impractical to manually look up the entire data and perform the taxonomy. Thus, we begin by randomly sampling 500 query pairs from the entire data and conducting manual classification. To validate our taxonomy, we randomly sample another 1,000 query pairs from the entire data as validation. We manually check whether the 1,000 query pairs can fit into the generated taxonomy. If not, we will include the new class in the existing ones and repeat the validation process. This process iterated until no new classes emerged, validating our taxonomy.

\textbf{Taxonomy result.} 
We taxonomize them into 8 categories (C1 to C8) summarized in Table~\ref{tab:taxonomy}. For each category, we provide a case study including the query pairs, where we search Query 1 and Query 2 on one of leading e-commerce retailers and compare the top 3 items. The analysis is as follows.

\begin{table*}[t]
\tabcolsep 0.2in
    \caption{Taxonomy and case study of robustness issues in e-commerce ranking system: The table classifies various types of query pairs and includes a case study with query pairs from one of leading e-commerce retailers.}
    \vspace{-0.2cm}
    \centering
    \begin{tabular}{cccc}

    \toprule
    & & \multicolumn{2}{c}{Case study}\\
    \cmidrule(lr){3-4}
    \multirow{-2}{*}{ID} & \multirow{-2}{*}{Taxonomy} & Query 1 & Query 2 \\
    \midrule
      C1&   Preposition & purple dress for women   &  women purple dress\\
    \multirow{-1}{*}{C2} & \multirow{-1}{*}{Abbreviation} & \multirow{-1}{*}{30'' marble top} & \multirow{-1}{*}{30 inch marble top} \\

    \multirow{-1}{*}{C3} & \multirow{-1}{*}{Singular and plural} & \multirow{-1}{*}{electric thing for kids} & \multirow{-1}{*}{electric things for kids} \\
    \multirow{-1}{*}{C4} &\multirow{-1}{*}{Word order} & \multirow{-1}{*}{red watch} & \multirow{-1}{*}{watch red} \\
    \multirow{-1}{*}{C5} & \multirow{-1}{*}{Article} & \multirow{-1}{*}{heels} & \multirow{-1}{*}{the heels} \\
    \multirow{-1}{*}{C6} & \multirow{-1}{*}{Punctuation} & \multirow{-1}{*}{funding} & \multirow{-1}{*}{funding.}\\
    \multirow{-1}{*}{C7}& \multirow{-1}{*}{Space} &\multirow{-1}{*}{24 x 20 outdoor cushion}  & \multirow{-1}{*}{24x20 outdoor cushion}\\

    \multirow{-1}{*}{C8} & \multirow{-1}{*}{Words connection} & \multirow{-1}{*}{black swing coat} & \multirow{-1}{*}{black+swing+coat} \\

    \bottomrule
         
    \end{tabular}

    \label{tab:taxonomy}
    \vspace{-0.2cm}
\end{table*}

\begin{itemize}[leftmargin=0.4cm]
  \item C1: Rewriting phrases using prepositions with semantic integrity confuses the system such as `T-shirt for men' vs. `men T-shirt'.
  \item C2: The use of abbreviations such as `volt' to `v' or `inch' to `in' confuses the ranking algorithm.
  \item C3: Switching between singular noun and plural noun, e.g., `T-shirt for man' vs. `T-shirts for men', leads to divergent results.
  \item C4: Queries where changing word order doesn't alter the semantics but affects rankings, such as `battery AA' vs. `AA battery'.
  \item C5: Including articles such as `the' in `the heels' produces entirely different ranking lists, especially related to movies and books.
  \item  C6: Similar to articles, punctuations mainly affect book and movie titles and thus the ranking results.
  \item  C7: Different treatment of spaces, e.g., `1 mm ring' vs. `1mm ring', leads to ranking differences.
  \item  C8: User or bot-generated queries might use different characters such as `+' or `x' to link words, affecting the ranking results.
\end{itemize}

\begin{tcolorbox}[boxrule=0pt, left = 2mm, right = 2mm, top = 0.5mm, bottom = 0.5mm, title= Observation \stepcounter{countobs}\arabic{countobs}]
Our taxonomy categorizes semantically identical query pairs challenging the e-commerce ranking system's robustness into eight classes, highlighting areas needing improvement.
\end{tcolorbox}
\section{Solutions and Future Directions}
\label{sec:solution}

\begin{figure}[!t]
\centering
\includegraphics[width=\linewidth]{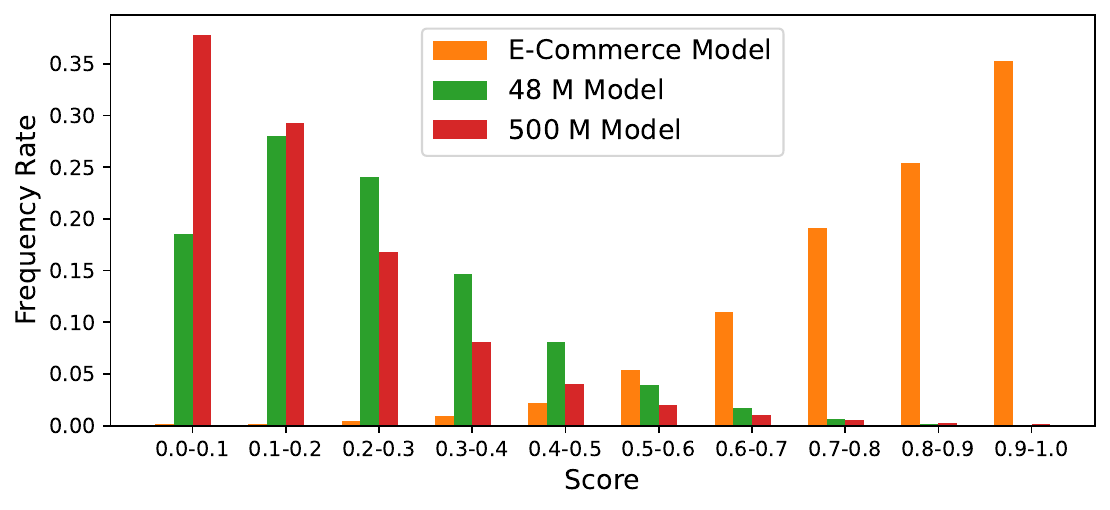}
\vspace{-0.8cm}
\caption{Comparison of histograms between the original e-commerce ranking model and two larger models on millions of \tps query pairs. Lower scores indicate greater robustness.}
\vspace{-0.5cm}
\label{fig:llm}
\end{figure}

\begin{table}[!t]
\tabcolsep 0.03in
\renewcommand{\arraystretch}{0.9}
    \caption{Evaluation of semantically identical query pairs using LLMs on various robustness issues in Table~\ref{tab:taxonomy}. Each class has 50 query pairs, and the LLMs answer `Yes' or `No'.}
    \vspace{-0.2cm}
    \centering
    \begin{tabular}{cccccccccc}

    \toprule
    LLM & C1 & C2 & C3 & C4 & C5 & C6 & C7 & C8 & Average \\
    \midrule
    GPT-3.5 & 76\% & 88\% & 36\% & 70\% & 68\% & 92\% & 98\% & 88\% & 77.0\%\\
    GPT-4.0 & 98\% & 100\% & 100\% & 100\% & 100\% & 100\% & 100\% & 98\% & 99.5\%\\
    
    \bottomrule
         
    \end{tabular}

    \label{tab:gpt}
    \vspace{-0.2cm}
\end{table}

\subsection{S1: Large Language Models}
\label{sec:s1}
As shown in Table~\ref{tab:taxonomy}, our analysis reveals that the prevalent issues in current e-commerce ranking models largely arise from their limited semantic understanding of user queries.
This limitation often leads these models to incorrectly differentiate between semantically identical query pairs. This, thus, results in significant ranking discrepancies as noted in our Observations 1 and 3. Thus, a promising improvement method for addressing these shortcomings involves the integration of Large Language Models (LLMs)~\cite{koubaa2023gpt}. These models excel in capturing both the semantic and syntactic nuances of queries, compared to the existing e-commerce ranking.

To substantiate this claim, we conduct preliminary tests using GPT-3.5 and GPT-4.0~\cite{koubaa2023gpt}. We randomly sampled 50 semantically identical query pairs from each taxonomy in~\S~\ref{sec:rq5}, each of which received disparate rankings from the existing e-commerce system. The LLMs are then tasked with determining whether the pairs are indeed semantically identical. The outcomes are summarized in Table~\ref{tab:gpt}. Notably, GPT-4.0 achieved a 99.5\% average accuracy in correctly identifying these pairs, while GPT-3.5 can provide 77.0\% accuracy on average. Thereby, LLMs potentially enhance the query understanding component of current e-commerce ranking systems.

Preliminary tests with GPT models indicate the potential benefits of using LLMs for improving ranking robustness. However, the impact of these improvements in an end-to-end setting remains unclear. Thus, we conduct additional experiments on two in-house ranking models, comprising billions of data points. We then compare their performance with that of a current e-commerce ranking model (\S\ref{sec:eval}), which is less complex than the two in-house models.

{\bf Methodology and setup.}
We select two representative in-house language ranking models with sizes of 48 million and 500 million parameters, respectively. 
For evaluation, we select at least 2 million query pairs, including all types in Table~\ref{tab:taxonomy}. Our measurement methodology and experimental setup align with those in~\S\ref{sec:methodandsetup}.
Note that we do not differentiate among various categories during the evaluation due to the high labor costs. Instead, we conduct large-scale testing involving millions of queries, rendering our results comparable to those of e-commerce ranking systems in~\S\ref{sec:eval}.

{\bf Result.} 
The comparison results on these three models are shown in Fig.~\ref{fig:llm}.
The histogram clearly indicates that both large-parameter ranking models demonstrate enhanced robustness in comparison to the current e-commerce model. Notably, the model comprising 500M parameters shows approximately twice as many queries falling within the 0.0 - 0.1 RDS range (0.0: a perfect match) as compared to its 48M counterpart. In contrast, the e-commerce ranking model scarcely exhibits cases within this optimal range. Additionally, the e-commerce model manifests a high frequency of queries with RDS exceeding 0.9 (1.0: no match between two ranked lists), which is scarcely observed in the two large-parameter models. Thus, large models offer significant potential for enhancing robustness.

While the current LLM results show effectiveness in addressing robustness issues, it is crucial to consider the latency requirements for services, such as requiring millisecond-level responsiveness~\cite{amazonlatency}. LLMs lead to substantial computational costs, necessitating a plethora of machines and GPUs~\cite{chatgptgpu}. Such overhead may be prohibitive for small companies due to budget constraints.

\begin{tcolorbox}[boxrule=0pt, left = 2mm, right = 2mm, top = 0.5mm, bottom = 0.5mm, title= Observation \stepcounter{countobs}\arabic{countobs}]
Large Language Models offer the potential to improve the robustness of e-commerce ranking systems but their high computational and financial costs constrain universal applicability.
\end{tcolorbox}

\subsection{S2: Model Ensemble}
\label{sec:s2}
We integrate model ensemble~\cite{yang2021certified} during inference by leveraging a few ranking models. The objective of this approach is to refine prediction quality by using aggregation techniques such as majority voting~\cite{parhami1994voting}. In the context of e-commerce ranking systems, the final output can be determined by averaging the positions across various ranking lists. To validate this hypothesis, we conduct an evaluation.

We assume that the e-commerce ranking system undergoes monthly updates. To evaluate the system's performance over time, we use 1,000 semantically identical query pairs that persist across a five-month period, specifically from April to August 2023. Subsequently, we aggregate the results over this time to calculate the smoothed RDS and compare it with the RDS for a single month. This assesses the efficacy of the model ensemble during inference.

The detailed histogram of RDS is presented in Table~\ref{tab:smoothing}. Specifically, in the most favorable case (i.e., a RDS range of 0.0--0.2), the rate of query pairs with model ensemble is 4.4 times higher than the rate observed without model ensemble. Conversely, in the least favorable case, the rate of query pairs with model ensemble is 1.2 times lower than that without model ensemble. These findings suggest that model ensemble enhances the robustness of the e-commerce ranking system during the inference stage, constituting a promising avenue for future research.

\begin{tcolorbox}[boxrule=0pt, left = 2mm, right = 2mm, top = 0.5mm, bottom = 0.5mm, title= Observation \stepcounter{countobs}\arabic{countobs}]
Model ensemble serves as a cost-effective method to improve the robustness of e-commerce ranking systems without requiring model training, offering a promising future direction.
\end{tcolorbox}

\begin{table}[t]
\renewcommand{\arraystretch}{0.9}
    \caption{RDS histogram with and without ensemble.}
    \vspace{-0.2cm}
    \centering
    \begin{tabular}{cccccc}

    \toprule
     & 0.0-0.2 & 0.2-0.4 & 0.4-0.6 & 0.6-0.8 & 0.8-0.1.0  \\
    \midrule
    No ensemble & 0.5\% & 2.5\% & 10.0\% & 30.3\% & 57.7\%\\
    Ensemble & 2.2\% & 4.7\% & 13.9\% & 32.1\% & 47.1\%\\
    
    \bottomrule
         
    \end{tabular}

    \label{tab:smoothing}
    \vspace{-0.3cm}
\end{table}

\subsection{S3: User Behavior Features Improvement}
\label{sec:s3}

Based on the information in Table~\ref{tab:taxonomy}, different semantically identical queries can significantly impact the ranking outcomes in e-commerce systems with ML algorithms. Essentially, these variations induce disparities in the feature sets used for ranking, thus, altering the model's output~\cite{kim2021distilling}. State-of-the-art e-commerce ranking systems are heavily dependent on features derived from user behavior. However, these features are prone to various forms of bias. For example, users are constrained to select items ranked by the e-commerce system, which generally relies on historical user behavior data for training. Thus, this can inadvertently introduce position bias into the new model. A ranking system that lacks initial robustness risks entering a self-perpetuating cycle of non-robustness, where flawed user behavior data and ranking outcomes mutually reinforce each other's weaknesses.
Therefore, introducing robustness during the training phase by refining the feature set can break the ill cycle and thus lead to significant improvements in the model's performance. While the details of our e-commerce ranking features cannot be disclosed due to legal constraints, we can discuss potential future directions.

Leveraging our framework and metrics, we can systematically and automatically identify query pairs that exhibit non-robust behavior. Armed with this, we can conduct controlled experiments by inputting these pairs into the ranking system and analyzing the feature differences. This is crucial for worst-case scenarios (RDS is 1.0) that may necessitate immediate remedial action in productions. Such experiments can elucidate which features are susceptible to robustness issues. The findings can guide the model training to improve robustness and accuracy. 
Additionally, our RDS could be integrated into training as a penalty, which is future research.

\begin{tcolorbox}[boxrule=0pt, left = 2mm, right = 2mm, top = 0.5mm, bottom = 0.5mm, title= Observation \stepcounter{countobs}\arabic{countobs}]
Our findings offer valuable insights for enhancing feature quality and interpretability in e-commerce ranking models, particularly in addressing worst-case predictive scenarios.
\end{tcolorbox}

\subsection{S4: DNN Robustness Improvement}
\label{sec:s4}
The ongoing tug-of-war between
adversarial attacks~\cite{carlini2017towards, zhang2020interpretable} and their defenses~\cite{xu2017feature, li2023sok} has yielded many robustness improvement strategies, such as adversarial training~\cite{bai2021recent}. Thus, we review the existing works to identify potential robustness improvement methodology that could enhance e-commerce ranking system robustness.

We categorize existing robustness improvement methods into two classes~\cite{cao2021invisible}: 1) model-level robustness improvements; and 2) input- or target-level robustness improvements. To the best of our knowledge, the majority of existing research focuses on 2), owing to its lightweight nature. These methods often include techniques such as automatic grammar checking~\cite{liu2022order}, spam detection~\cite{liu2022order, chen2023defense}, and frequency-guided word substitutions~\cite{mozes2020frequency}. Notably, these methodologies are incorporated into commercial e-commerce systems. Our empirical analyses are based on evaluations conducted within a commercial environment, leading to the following observation.

\begin{tcolorbox}[boxrule=0pt, left = 2mm, right = 2mm, top = 0.5mm, bottom = 0.5mm, title= Observation \stepcounter{countobs}\arabic{countobs}]
While input- or target-level robustness improvement is commonly used in e-commerce systems, they fall short of effectively improving the robustness of the ranking models.
\end{tcolorbox}

Due to the paucity of research on model-level robustness in ranking systems, we perform a comprehensive literature review in other domains such as image and NLP. Certified robustness~\cite{li2023sok} and adversarial training~\cite{goibert2019adversarial} are widely acknowledged as potent methodologies for enhancing model robustness. However, these approaches have not yet been explored within the sphere of ranking systems, especially those in e-commerce systems.

Adversarial training, which aims to improve model robustness by solving a min-max optimization problem, often adversely affects the performance of the normal or benign model~\cite{zhang2019theoretically,wang2018kdgan,wang2019adversarial}. Such a trade-off is typically unacceptable in commercial settings. Besides, the computational overhead associated with adversarial training is substantial~\cite{sriramanan2021towards}, rendering it economically unfeasible for many organizations. Thus, improving the efficiency and performance of e-commerce ranking systems is an important future work.

Regarding certified robustness, although it offers theoretical guarantees of robustness, its current applications are limited to small models and simpler tasks, such as image classification~\cite{li2023sok}. This limitation is primarily due to the complexity of the methodology and the associated computational costs. Thereby, these potent robustness-improving methods are conspicuously absent in the current research on ranking models, which requires future efforts.

\begin{tcolorbox}[boxrule=0pt, left = 2mm, right = 2mm, top = 0.5mm, bottom = 0.5mm, title= Observation \stepcounter{countobs}\arabic{countobs}]
To the best of our knowledge, model-level robustness improvement in ranking models remains underexplored. Due to their potential for robustness enhancement and theoretical guarantees, these approaches offer a compelling future direction.
\end{tcolorbox}

\section{Discussion}
\label{sec:discussion}

\begin{figure}[!t]
\centering
\includegraphics[width=0.95\linewidth]{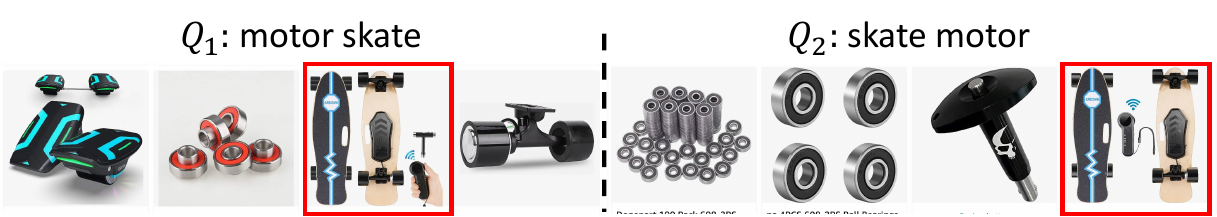}
\vspace{-0.2cm}
\caption{Example of two semantically distinct queries sharing a same ranked item on the e-commerce retailer.}
\label{fig:sematic_diff}
\vspace{-0.4cm}
\end{figure}

\begin{figure}[!t]
\centering
\includegraphics[width=0.95\linewidth]{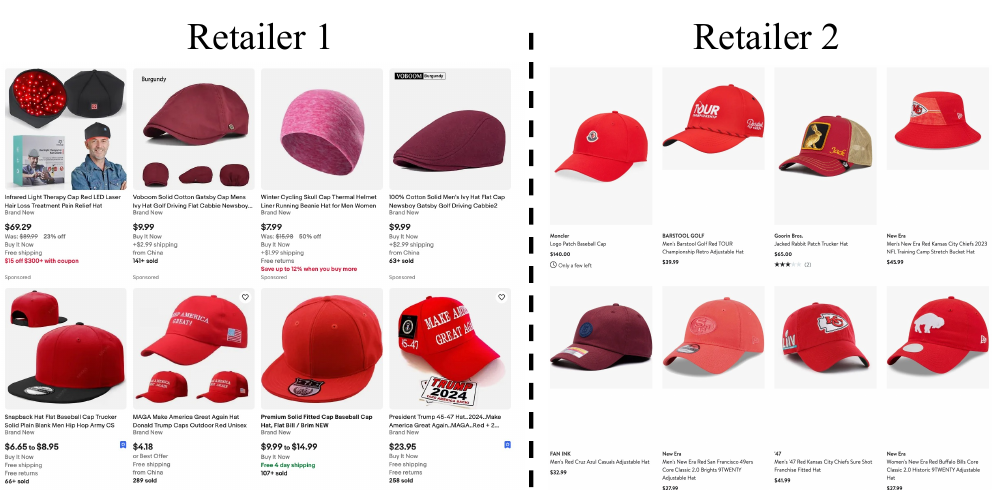}
\vspace{-0.3cm}
\caption{Search results for the query `hat not red' on two commercialized e-commerce retailers.}
\label{fig:hatnored}
\vspace{-0.4cm}
\end{figure}

{\bf Other interesting problems.} We observe two interesting types of issues: 1) incongruities in item ranking and 2) difficulties in handling negations. 
In the first category, the e-commerce ranking system displays a limited capacity for semantic discernment between queries. This lack of semantic understanding leads to overlapping items in the ranking lists for semantically distinct queries. For example, queries like `motor skate' and `skate motor' yielded a shared ranked item depicted in Fig.~\ref{fig:sematic_diff}. In the second category, \textit{negation} poses a substantial challenge. A query such as `hat not red' frequently results in the display of red hats, indicating a failure to understand the negation. Fig.~\ref{fig:hatnored} offers an example from two e-commerce retailers. These observations suggest that current e-commerce ranking systems have substantive issues, requiring further enhancement.

\textbf{Limitations.} First, its scope is confined to the U.S. market and English language on one e-commerce retailer. While its prominence and ubiquity lend a degree of generality to our findings, the results may not hold across different linguistic or cultural contexts. Second, there are potential biases that may affect the study, including data and ranking item bias. Despite efforts to mitigate these through data cleaning, residual biases may persist. Third, the ranking list is comprised of item hashing IDs, which are not semantically analyzed in this study. For instance, some items are very similar but use different IDs. Thus, future work could benefit from incorporating semantic understandings of these IDs into metrics and analyses.
\section{Conclusion}
\label{sec:conclusion}
In this paper, we perform a systematic exploration of the robustness of a leading commercialized e-commerce ranking system. We propose a novel metric for conducting large-scale measurement studies and presenting insights and solutions. Our investigation uncovers a general issue within e-commerce ranking systems: the lack of robustness in producing consistent ranking outcomes for semantically identical queries. We hope that our findings and insights can further inspire the research of e-commerce ranking systems to potently improve the robustness and accuracy.

\section{Acknowledgments}
We would like to thank Qi Alfred Chen, Vamsi Salaka, Abbas Kazemipour, Hancao Li, Zhongruo Wang, Zhen Zhang, Junze Liu, Dan Luo, Ziwen Wan, Xinyang Zhang, Tong Wu, and the anonymous reviewers for their valuable and insightful feedback.

\bibliographystyle{ACM-Reference-Format}
\bibliography{main}


\end{document}